\documentclass[aps,prx,twocolumn,floatfix,10pt,superscriptaddress]{revtex4-1}
\usepackage[utf8]{inputenc}

\usepackage{natbib}
\usepackage{graphicx}
\usepackage{latexsym}
\usepackage{amssymb}
\usepackage{amsmath}
\usepackage{amsfonts}
\usepackage{gensymb}
\usepackage{bm}
\usepackage{hyperref}
\usepackage{braket}
\usepackage{physics}
\usepackage{bbold}
\usepackage[caption=false]{subfig}
\usepackage[dvipsnames]{xcolor}
\usepackage[normalem]{ulem}
\usepackage{siunitx}
\usepackage{mhchem}

\hypersetup{
  pdfauthor = {Taige Wang},
  pdftitle = {}
}

\newcommand{\supplementarysection}{%
  \setcounter{figure}{0}
  \let\oldthefigure\thefigure
  \renewcommand{\thefigure}{S\oldthefigure}
  \setcounter{section}{0}
  \let\oldthesection\thesection
  \renewcommand{\thesection}{S\oldthesection}
  \setcounter{equation}{0}
  \let\oldtheequation\theequation
  \renewcommand{\theequation}{S\oldtheequation}
  \setcounter{table}{0}
  \let\oldthetable\thetable
  \renewcommand{\thetable}{S\oldthetable}
}

\begin{document}

\title{Diverse magnetic orders and quantum anomalous Hall effect\\ in twisted bilayer \ce{MoTe2} and \ce{WSe2}}

\author{Taige Wang}
\affiliation{Department of Physics, University of California, Berkeley, CA 94720, USA \looseness=-1}
\affiliation{Material Science Division, Lawrence Berkeley National Laboratory, Berkeley, CA 94720, USA \looseness=-1}

\author{Trithep Devakul}
\affiliation{Department of Physics, Massachusetts Institute of Technology, Cambridge, MA 02139 USA}

\author{Michael P. Zaletel}
\affiliation{Department of Physics, University of California, Berkeley, CA 94720, USA \looseness=-1}
\affiliation{Material Science Division, Lawrence Berkeley National Laboratory, Berkeley, CA 94720, USA \looseness=-1}

\author{Liang Fu}
\affiliation{Department of Physics, Massachusetts Institute of Technology, Cambridge, MA 02139 USA}

\date{\today}

\begin{abstract}
Twisted homobilayer transition metal dichalcogenide (TMD) offers a versatile platform for exploring band topology, interaction-driven phases, and magnetic orders.
We study the interaction-driven phases in twisted TMD homobilayers and their low-energy collective excitations, focusing on the effect of band topology on magnetism and its thermal stability.
From Hartree-Fock theory of the continuum model, we identify several magnetic and topological phases. By tuning the displacement field, we find two phase transitions involving a change in topology and magnetism respectively. 
We analyze the magnon spectrum, revealing the crucial role of band topology in stabilizing 2D ferromagnetism by amplifying easy-axis magnetic anisotropy, resulting in a large magnon gap of up to \SI{7}{meV}.
As the magnon gap is directly tied to the stability of the magnetic phase to thermal fluctuations, our findings have several important experimental implications.
\end{abstract}

\maketitle

{\em Introduction --} Harnessing the magnetic properties of materials is a long-standing goal in condensed matter physics.  
In conventional magnets, magnetism is associated with magnetic elements having partially filled $d$ and $f$ orbitals, which is difficult to control externally.  However, recent experimental breakthroughs in two dimensional moir\'e heterostructures have opened up new avenues for tailoring magnetism~\cite{andrei2021marvels,mak2022semiconductor}.  By tuning control parameters such as twist angle, electrostatic gating, or electric field, it becomes possible to significantly modify the moir\'e band structure and induce a wide range of interaction-driven states with magnetic orders and topological states. 
Such interaction-driven magnetism without magnetic elements can even be tuned {\it in situ} by electrical means. This unprecedented control of magnetism has potential in developing spintronic devices and quantum technologies.

For 2D materials, a practical question is the stability of magnetic phases against thermal fluctuations, which is governed by the nature of low-energy collective modes. Magnetic orders that break a continuous spin rotation symmetry, such as spin \( SU(2) \), necessarily give rise to gapless Goldstone modes. These modes generate fluctuations that prevent long-range magnetic order in 2D at any finite temperature, as dictated by the Mermin-Wagner theorem. Consequently, robust ferromagnetism in 2D requires strong magnetic anisotropy to overcome these fluctuations. In pristine graphene, where spin-orbit coupling is negligibly small, spin ferromagnetism cannot establish itself as a true long-range order at finite temperature. Even in monolayer transition metal dichalcogenides (TMDs) with strong Ising spin-orbit coupling, the locking of spin \( s_z = \uparrow (\downarrow) \) to valley \( K (K') \) leads to an emergent pseudospin \( SU(2) \) symmetry at low doping, which again suppresses long-range magnetic order. This challenge is finally overcome in moir\'e TMD homobilayers, where the combination of spin-valley locking and valley-dependent interlayer tunneling explicitly breaks the pseudospin symmetry down to \( U(1) \times \mathbb{Z}_2 \) \cite{WuPRL}. As a result, long-range Ising-type magnetic order at finite temperatures is made possible via spontaneous spin \( s_z \) polarization, which preserves \( U(1) \) symmetry while breaking \( \mathbb{Z}_2 \) spin symmetry \cite{LiangNC}. This phenomenon has been exemplified by recent observations of Ising ferromagnetism, characterized by clear magnetic hysteresis, in twisted homobilayer TMDs at small twist angles \cite{Xiaodong,XiaodongFQH,zeng2023integer}. Furthermore, when ferromagnetism is combined with non-trivial band topology in TMD bilayer systems, it can lead to a quantized anomalous Hall effect at integer fillings, a topic that has garnered significant interest~\cite{li2021quantum,Ben,wang2020correlated,LiangPRX,pan2022topological,xie2022valley,xie2022topological,dong2023excitonic,chang2022quantum,pan2020band,zang2021hartree,mai20221,xie2022nematic,Multiferroicity,patra2023electric,ryee2023switching,qiu2023interaction}.

In this work, we explore the rich phase space of magnetic orders in twisted TMD homobilayers and their low-energy collective excitations, with a particular focus on how band topology influences magnetism. Specifically, for twisted homobilayers of \ce{MoTe2} and \ce{WSe2} at twist angles in the range \(1^\circ < \theta < 4^\circ\), and under realistic interaction conditions, we map out the ground state phase diagram at the filling of one hole per moir\'e unit cell using self-consistent Hartree-Fock (HF) theory of the continuum model. As a function of twist angle and displacement field, we uncover a variety of magnetic and topological states, including valley polarized (VP) states and inter-valley coherent (IVC) states, which exhibit Ising and $XY$ magnetic orders, respectively. Furthermore, we calculate the spectrum of magnon excitations using the time-dependent HF (TDHF) formalism \cite{SoftMode,wu2020ferromagnetism,wu2020collective,kwan2021exciton}.

We show that robust Ising ferromagnetism is intimately connected to the presence of topological moir\'e bands, which are characterized by the spin (or equivalently, valley) Chern number. In particular, band topology stabilizes 2D ferromagnetism by amplifying the easy-axis magnetic anisotropy, leading to a sizable magnon gap of up to approximately \SI{7}{meV}. Upon applying a displacement field, we observe two distinct transitions within most of the examined twist angle range. First, the valley-polarized (VP) state undergoes a topological phase transition from a Chern insulator to a trivial insulator, driven by electronic band inversion, yet without affecting the underlying Ising ferromagnetism. One of our key findings is the significant decrease in the magnon gap across this transition, underscoring the critical role of band topology in enhancing magnetic anisotropy. As the displacement field is further increased, a second magnetic phase transition occurs, wherein the system shifts from an Ising ferromagnetic (VP) state to an $XY$ antiferromagnetic (IVC) state. This transition can be understood as the condensation of magnons at a commensurate wave vector, corresponding to a continuous magnetic canting transition, during which the charge gap remains unclosed.

\begin{figure}
    \centering
    \includegraphics[width = 0.44\textwidth]{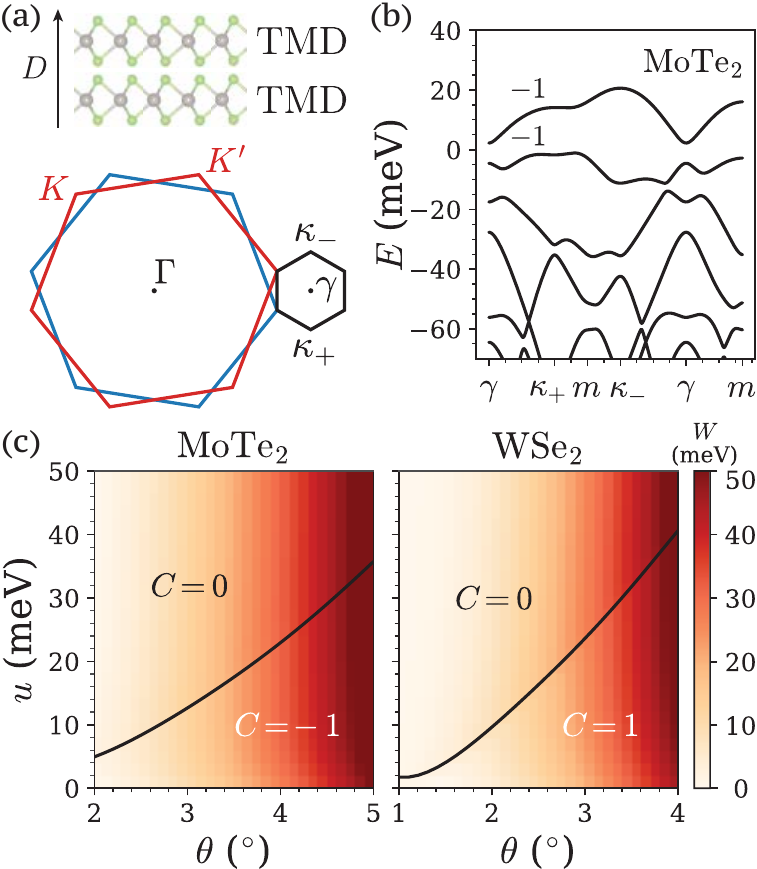}
    \caption{(a) Atomic structure and Brillouin zone (BZ) of twisted homobilayer TMD. The red and blue BZ are the atomic BZ of the top and the bottom layer respectively, and the black BZ is the resulting moir\'e BZ. (b) Typical band structure of twisted homobilayer TMD, with twisted \ce{MoTe2} with twist angle $\theta = 3.5 \degree$ and displacement field $u_D = \SI{5}{meV}$ as an example. We label the valley Chern numbers of the first two valence bands in the $K$ valley, which depend on both the twist angle and the displacement field. (c) Band width and valley Chern number of the first valence band in the $K$ valley as a function of twist angle and displacement field.}
    \label{fig:free}    
\end{figure}

{\em Continuum model --} In twisted TMDs, the valence band edge is located at the $K$ and $K'$ point of the Brillouin zone \cite{LiangNC,LiangFQH}. The strong Ising spin-orbit coupling locks the spin degree of freedom to the valley degree of freedom at low energy. Following Ref.~\onlinecite{WuPRL}, we write down the continuum model Hamiltonian,
\begin{equation} \label{eq:hamiltonian}
\begin{gathered}
    H_0=\sum_{\tau, l, \mathbf{r}} c_{\tau, l, \mathbf{r}}^{\dagger}\left([h_{\tau}]_{ll'}-\mu \delta_{l l'}\right) c_{\tau, l', \mathbf{r}}\\
    [h_{K}]_{ll'}=\left(\begin{array}{cc}h^K_+ + V_{+}(\boldsymbol{r}) - u_D & T(\boldsymbol{r}) \\ T^{\dagger}(\boldsymbol{r}) & h^K_- + V_{-}(\boldsymbol{r}) + u_D\end{array}\right)
\end{gathered}
\end{equation}
where $\tau$ is the valley index and $l$ is the layer index. Here $h^K_{\pm}$ is the kinetic part, $V_{\pm}(\mathbf{r})$ and $T(\mathbf{r})$ are the intralayer moir\'e potential and interlayer tunneling respectively due to the moir\'e structure. In the Supplementary Material \cite{SM}, we give the explicit form of $V_{\pm}(\mathbf{r})$ and $T(\mathbf{r})$, and the continuum model parameters from recent large-scale DFT calculations \cite{LiangNC,LiangFQH}. A vertical displacement field is modeled as a layer potential difference $u_D$.  The $K'$ valley hamiltonian $h_{K'}$ can be obtained as the time-reversal conjugate of $h_K$.

In additional to the moir\'e point group symmetries, i.e., the translation $T_{1,2}$ and the threefold rotation $C_3$ symmetry, the twisted TMD Hamiltonian in Eqn.~\ref{eq:hamiltonian} also respects the total charge $U(1)_c$, the valley charge $U(1)_v$, and the time reversal symmetry $\mathcal{T}$. We typically consider finite vertical displacement fields which breaks the $C_{2y}$ rotation symmetry at $u_D=0$.

The typical non-interacting band structure of twisted TMD is shown in Fig.~\ref{fig:free} (b). We calculate the bandwidth and the valley Chern number of the first valence band for twisted \ce{MoTe2} and \ce{WSe2} at various twist angles and displacement fields, shown in Fig.~\ref{fig:free} (c). 
At zero displacement field, twisted \ce{WSe2} has valley Chern number $C_K = 1$ and twisted \ce{MoTe2} has $C_K = -1$ \cite{LiangFQH}. The valley Chern number vanishes beyond a critical displacement field $u_D^{c0}$, which depends on the twist angle. As we will now show, the twist angle and the tunable valley Chern number have a profound influence on the interaction-driven phases.

{\em Interaction-driven phases --} Now we examine the effect of interaction by adding a gate-screened Coulomb interaction to the Hamiltonian. We study the interacting Hamiltonian with band-projected self-consistent Hartree-Fock numerics, as detailed in the Supplementary Material \cite{SM}. As shown in Fig.~\ref{fig:pd}, we observe a rich interacting phase diagram at one hole per moir\'e unit cell that features various magnetic orders. For realistic dielectric constant $\epsilon < 25$, we always observe an incompressible state that breaks various combinations of the valley $U(1)_v$ symmetry and the time reversal $\mathcal{T}$ symmetry~\cite{SM}. Here we restrict ourselves to two \textit{active} bands per valley, since the third band can affect parts of the phase diagram of twisted \ce{MoTe2} in a way which sensitively depends on the phase of the moir\'e potential (see Supplementary Material for a detailed discussion).

\begin{figure}
    \centering
    \includegraphics[width = 0.48\textwidth]{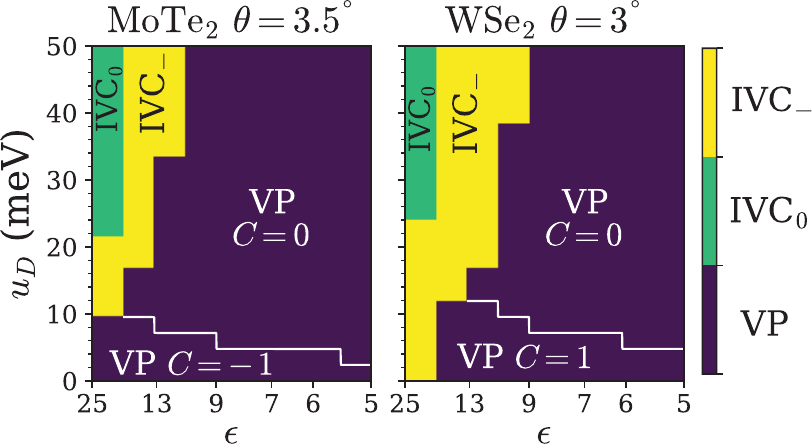}
    \caption{(a) Interacting phase diagram of twisted \ce{MoTe2} and \ce{WSe2} as a function of displacement field and dielectric constant. Different colors represent three different phases: VP, IVC$_{\mathbf{0}}$, and IVC$_-$ (see main text for definition). We label the regime where the ground state has non-zero Chern number.}
    \label{fig:pd}    
\end{figure}

The prominent states in play are the valley polarized (VP) state, inter-valley coherent (IVC) state at zero momentum (IVC$_{\mathbf{0}}$), and IVC state at momentum $\mathbf{Q} = \kappa_+ - \kappa_- = \kappa_-$ (IVC$_-$) (see Fig.~\ref{fig:free} (a)) \footnote{The moir\'e scale momentum $\mathbf{q}$ is defined relative to $\gamma_K - \gamma_{K'}$, the momentum transfer between the $\gamma$ point of two valleys, which is an integer multiple of the moir\'e reciprocal lattice vector $\mathbf{G}$ if the moir\'e unit cell is commensurate \cite{IKS,IKSSTM}.}. Due to spin-valley locking, the VP state is an Ising ferromagnet with out-of-plane spin polarization. The IVC states show three-sublattice $120^\circ$ antiferromagnetic (AFM) order at atomic scale. The order parameter can be represented as a complex scalar field $S^+ \equiv  S_x + i S_y$ and we label its moir\'e scale momentum $\mathbf{q}\ll K$,
\begin{equation}
    S^+_{\mathrm{IVC}}(\mathbf{q})=\sum_{\mathbf{r}} e^{-i\left(\mathbf{K}^{\prime}-\mathbf{K}+\mathbf{q}\right) \cdot \mathbf{r}} S^+(\mathbf{r}) \sim \sum_{\mathbf{k}} \langle \psi_{K, \mathbf{k}}^{\dagger} \psi_{K', \mathbf{k}+\mathbf{q}} \rangle
\end{equation}
where $\psi_{\tau, \mathbf{k}}^{\dagger}$ denotes the moir\'e Bloch electrons of the $\tau$ valley. 
The magnetic orders of various valley-ordered states are summarized in Table~\ref{tab:correspondence} \footnote{All in-plane magnetic orders are only meaningful at the level of effective Kane-Mele model at moir\'e scale because of the atomic scale $120^{\circ}$ AFM order.}.

\begin{table}[!htbp]
    \centering
    \begin{tabular}{cc}
    \hline \hline Magnetic orders & Valley-ordered phases \\
    \hline QAH & topological VP\\    FM$_z$ & trivial VP\\
    FM$_{xy}$ & IVC$_{\mathbf{0}}$\\
    $120 \degree$ AFM & IVC$_-$\\
    \hline \hline
    \end{tabular}
    \caption{Magnetism and topology of various valley-ordered states.}
    \label{tab:correspondence}
\end{table}

\begin{figure}
    \centering
    \includegraphics[width = 0.48\textwidth]{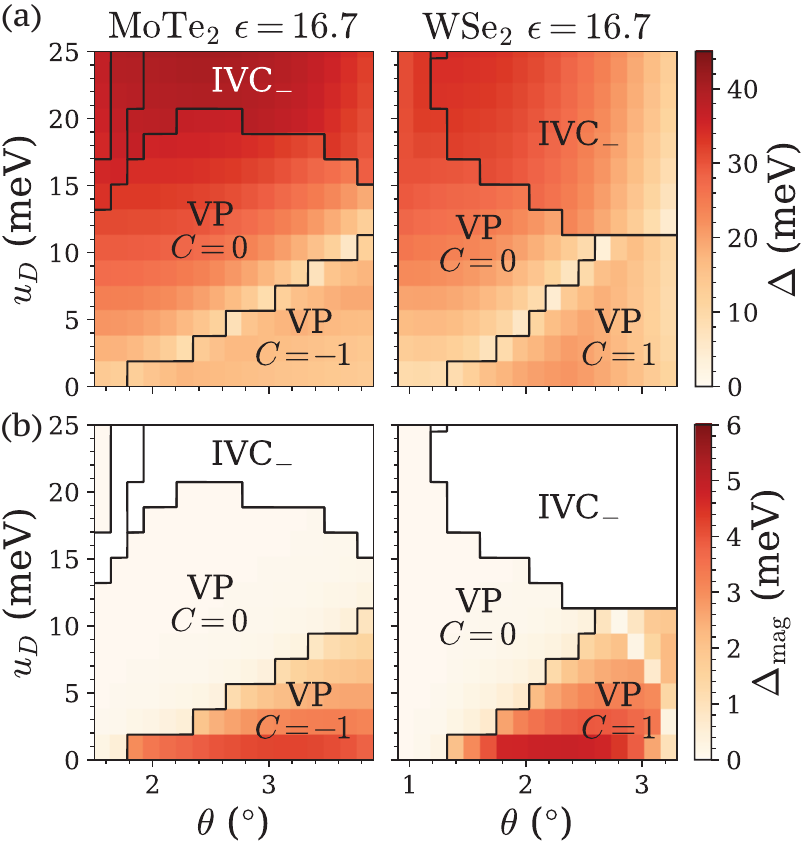}
    \caption{Interacting phase diagram of twisted \ce{MoTe2} and \ce{WSe2} as a function of displacement field and twist angle. We show the charge gap $\Delta$ in (a) and the magnon gap $\Delta_{\mathrm{mag}}$ in (b). The IVC phase has gapless Goldstone modes instead of magnons, which is left blank in (b). We note that the magnon gap is non-vanishing, but parametrically small in the regime where the VP state is topologically trivial.}
    \label{fig:gap}    
\end{figure}

Both topological and non-topological VP phases exist in the phase diagram of Fig.~\ref{fig:pd}.
The topological trivial VP phase is adiabatically connected to the multiferroic phase discussed in Ref.~\onlinecite{Multiferroicity}, which persists down to zero displacement field at small twist angles (see Fig.~\ref{fig:gap} (a)). The IVC state can be viewed as a coherent superposition of the first valence bands of the two valleys. When these two bands possess nonzero and opposite Chern numbers, the phase of the IVC order parameter necessarily exhibits non-trivial winding in momentum space. This results in an additional interaction energy cost compared to the VP state \cite{BCZ2020,DGGmoire,Serlin2019,ZMS2019,NickPRX,TaigeNC}, therefore, the IVC phase shrinks at smaller dielectric constants (see Fig.~\ref{fig:pd}).

The displacement field is the most experimentally accessible tuning parameter, governing both the topological transitions and the bandwidth at the single-particle level. At small displacement fields, the large gap between the first two valence bands prevents interactions from mixing them. As a result, it is energetically favorable to fully occupy the first valence band in one valley, leading to a VP state with Chern number \( C = \pm 1 \), inherited from the non-interacting band structure. As the displacement field increases to a critical value \( u_D^c \), the bulk gap closes, signaling a topological phase transition from a Chern insulator to a topologically trivial VP state. At displacement fields well beyond \( u_D^c \), a second transition occurs, where the VP state gradually cants into an IVC state on the valley Bloch sphere. This canting transition occurs without closing the bulk charge gap \cite{MikePRX}.

In Fig.~\ref{fig:gap}, we present the phase diagram at a fixed dielectric constant \( \epsilon = 16.7 \) while varying the twist angle and displacement field. The charge gap shown can be directly compared to transport experiments. As the twist angle increases, the critical displacement field \( u_D^c \) for the topological transition also increases, following the trend of the non-interacting value \( u_D^{c0} \). In general, interactions tend to increase the layer polarization, resulting in a topologically trivial band. Conversely, as the twist angle increases, the IVC state occurs at a reduced \( u_D \), as it benefits from more dispersive bands at larger twist angles, achieved by canting the valley isospin vector in momentum space \cite{AdrianTBG,EslamNC,TaigeNC}. 

\begin{figure}
    \centering
    \includegraphics[width = 0.44\textwidth]{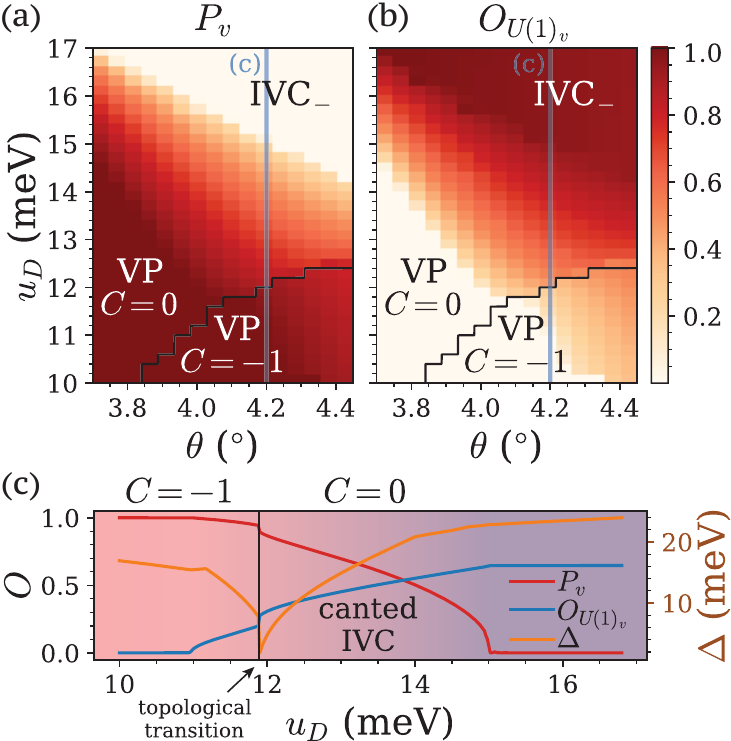}
    \caption{(a) Valley polarization and (b) \( U(1)_v \) symmetry breaking order parameter across the canting transition in twisted \ce{MoTe2} with dielectric constant \( \epsilon = 16.7 \) (see main text for definition). (c) A line cut at \( \theta = 4.2^\circ \). The charge gap \( \Delta \) vanishes at the topological transition.}
    \label{fig:cant}    
\end{figure}

At sufficiently large twist angles, the intermediate trivial VP phase is expected to disappear, and the topological and magnetic transitions merge into a single transition. We explore this parameter regime in Fig.~\ref{fig:cant}, where the interplay of these transitions is examined. Two key order parameters are shown: the valley polarization \( P_v \equiv (n_K - n_{K'})/(n_K + n_{K'}) \), which quantifies the relative occupation of the two valleys, and the \( U(1)_v \)-symmetry-breaking order parameter \( O_{U(1)_v} \equiv \sum_{\mathbf{k}} |\langle \psi_{K,\mathbf{k}}^{\dagger} \psi_{K',\mathbf{k}} \rangle | \), which measures the degree of inter-valley coherence. \( n_\tau \sim \sum_{\mathbf{k}} \langle \psi_{\tau,\mathbf{k}}^{\dagger} \psi_{\tau,\mathbf{k}} \rangle \) denotes the occupation of each valley.

At larger twist angles, such as \( \theta > 4^\circ \), the valley polarization \( P_v \) smoothly decreases, while the \( U(1)_v \)-symmetry-breaking order parameter \( O_{U(1)_v} \) increases over a displacement field range of approximately \SI{4}{meV}, as shown in Fig.~\ref{fig:cant}(b). The topological transition occurs within this canted regime, where both valley \( \mathbb{Z}_2 \) and \( U(1)_v \) symmetries are broken. The transition is continuous, as indicated by the vanishing charge gap. This canting-induced topological transition between \( C = -1 \) and \( C = 0 \) magnetic states can be described within the framework of charge transfer gap inversion theory~\cite{ZhenNC,LiangPRX}.

\begin{figure}
    \centering
    \includegraphics[width = 0.48\textwidth]{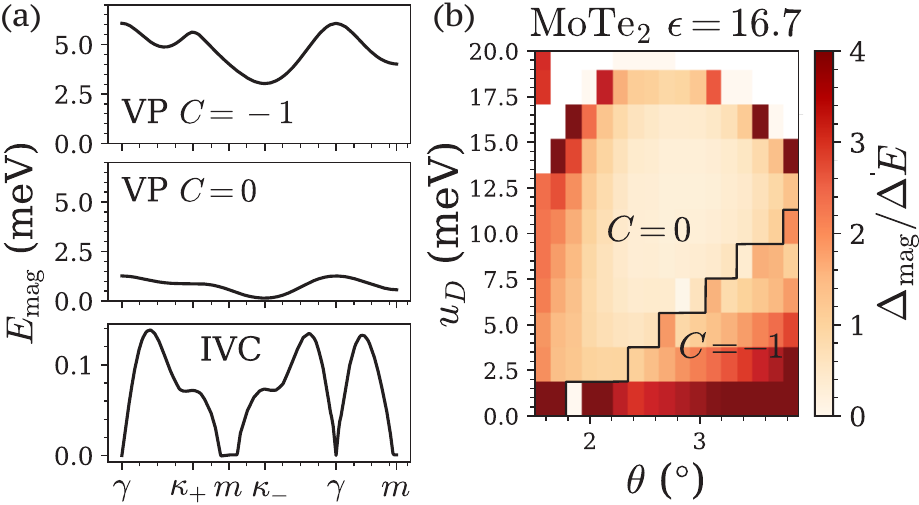}
    \caption{(a) Typical TDHF soft mode band structure of twisted TMD in the topological VP, trivial VP and the IVC phase. We take twisted \ce{MoTe2} at $\theta = 3.5^{\circ}$ with $\epsilon = 16.7$ at $u_D = \SI{2.5}{meV}$ (topological VP), \SI{11}{meV} (trivial VP), and \SI{25}{meV} (IVC) as an example. (b) Ratio of the energy difference between the VP and IVC state and the magnon gap.}
    \label{fig:magnon}    
\end{figure}

{\em Magnon and thermal stability of the magnetic orders --} Now we discuss the finite-temperature stability of these magnetic orders, which is controlled by the low energy collective modes. More precisely, thermal fluctuation of these modes, present at temperatures higher than their gap, leads to melting of the magnetic order. We employ the time-dependent HF (TDHF) calculation \cite{wu2020ferromagnetism,wu2020collective,kwan2021exciton}, as implemented for moir\'e heterostructures in Ref.~\onlinecite{SoftMode}, to compute all soft mode excitations in the HF ground state. 

We show the magnon band of twisted \ce{MoTe2} at $\theta = 3.5 \degree$ in Fig.~\ref{fig:magnon} as an example. In the topological VP phase, the magnon band is relatively flat and features a large magnon gap. 
Once we cross the transition into the trivial VP phase, the magnon gap drops significantly despite a similar magnon bandwidth. Here, we see the prototypical (almost) gapless quadratic Goldstone modes of a (nearly) isotropic ferromagnet (the pseudospin $SU(2)$ rotation symmetry becomes exact in our model when $u_D\rightarrow\infty$, which reduces to a monolayer with a periodic potential). When we finally reach the IVC$_-$ phase, the low energy soft modes are no longer magnons that carry spin quantum number. Instead, we find linearly dispersed gapless goldstone modes from spontaneously breaking of the valley $U(1)_v$ symmetry.

Fig.~\ref{fig:gap} (b) emphasizes the correlation between the non-trivial band topology and the magnon gap. The magnon gap can reach up to $\sim \SI{7}{meV}$ in the topological VP phase, while it quickly drops to almost zero across the topological transition. This behavior requires understanding the nature of the low-energy soft modes. In the VP phase (topological or not), we find that the soft mode at the magnon band bottom corresponds to a spin-valley flip with momentum moir\'e momentum $\mathbf{Q} = \kappa_-$, 
\begin{equation}
    \ket{m(\mathbf{Q})} \sim \sum_{\mathbf{k}} \psi_{K, \mathbf{k}}^{\dagger} \psi_{K', \mathbf{k}+\mathbf{Q}} \ket{\mathrm{VP}}
\end{equation}
We note that condensing such a magnon will result in an IVC state with momentum $\mathbf{Q}$, which is characterized by $S^+_{\mathrm{IVC}} \sim \sum_{\mathbf{k}} \langle \psi_{K, \mathbf{k}}^{\dagger} \psi_{K', \mathbf{k}+\mathbf{q}} \rangle$. In other words, IVC states can also be viewed as magnon superfluid. When the parent state is a topological VP state, the additional energy required to condense into the IVC state, or the magnon gap, is higher due to the aforementioned energy penalty against inter-Chern states.

The competition between the VP and IVC states, as well as the magnon dispersion, can be captured by the following non-linear sigma model (see the Supplementary Materials for a detailed discussion)\cite{SoftMode},
\begin{equation}
    \label{eq:nlsm}
\begin{aligned}
    \mathcal{L}[\tilde{Q}]=&\frac{1}{2} \operatorname{tr} T(\boldsymbol{r})^{\dagger} \tilde{Q} \partial_t T(\boldsymbol{r})-\frac{\rho}{4} \operatorname{tr}[\nabla \tilde{Q}(\boldsymbol{r})]^2 + \\&\frac{\alpha}{2} \operatorname{tr}\left(\tilde{Q} \tau_z\right)^2 - \frac{J}{4} \operatorname{tr}\left[\left(\tilde{Q} \tau_x\right)^2+\left(\tilde{Q} \tau_y\right)^2\right]
\end{aligned}
\end{equation}
where \( \tilde{Q}_{\alpha \beta} \equiv \mathbf{n} \cdot \boldsymbol{\tau} \) characterizes the polarization \footnote{The actual order parameter $Q_{\boldsymbol{k}}=e^{i \varphi(\boldsymbol{k}) \gamma_z} \tilde{Q} e^{-i \varphi(\boldsymbol{k}) \gamma_z}$ necessarily winds $4\pi$ around the Brillouin zone in the inter-Chern IVC phase.}, and the spatial variation of \( Q(\boldsymbol{r}) = T^{\dagger}(\boldsymbol{r}) Q T(\boldsymbol{r}) \) describes the magnon modes. Here, \( \rho \) is the stiffness of the magnon modes, \( \alpha \) represents the energy penalty against inter-Chern states due to order parameter winding, and \( J \) characterizes the AFM coupling from the dispersion.

Within this non-linear sigma model, the VP state is described by \( \tilde{Q}_{\mathrm{VP}} = \tau_z \), and the IVC state by \( \tilde{Q}_{\mathrm{IVC}} = \cos \phi \tau_x + \sin \phi \tau_y \), with their energy difference given by \( \Delta E = E_{\mathrm{IVC}} - E_{\mathrm{VP}} = \alpha - J/2 \). The magnon dispersion on top of the VP ground state can then be solved as
\begin{equation}
    \hbar \omega_{\mathbf{q}} = \rho q^2 + \alpha - \frac{J}2
\end{equation}
This equation shows that the magnon gap \( \Delta_{\mathrm{mag}} \) is approximately the energy difference \( \Delta E \) between the VP and IVC states. Although the order parameter \( Q \) has complex momentum-space structures, making Eqn.~\ref{eq:nlsm} an approximation, \( \Delta_{\mathrm{mag}} \) remains roughly-approximated by \( \Delta E \), as demonstrated in Fig.~\ref{fig:magnon}(b). The discrepancy primarily arises from the \( J \)-term, where the HF reduces energy by explicitly canting the isospin vector in momentum space. In contrast, in the coarse-grained nonlinear sigma model, the \( J \)-term is introduced by integrating out charge fluctuations \cite{skyrmion}.

Since the ferromagnetic ordering temperature is governed by \( \Delta_{\mathrm{mag}} \), the energy penalty \( \alpha \) against inter-Chern states plays a crucial role in stabilizing ferromagnetism at finite temperatures. Additionally, the nontrivial topology causes the electron wave function to spread out more, enhancing the exchange interaction that depends on wave function overlap \cite{QHFM}. The combined effects of increased magnetic anisotropy and enhanced exchange interaction bolster Ising ferromagnetism when the underlying electronic band is topologically non-trivial.


{\em Discussion --} Our results show that the presence of band topology significantly enhances the magnetic anisotropy, which in turn increases the magnon gap and Curie temperature. This has important implications for previous studies. Interaction-driven phase diagrams of TMD homobilayers from earlier works~\cite{pan2020band,zang2021hartree,LiangNC,Multiferroicity,qiu2023interaction}, which employed various theoretical techniques and models, have found similar magnetic and topological phase candidates, including topological and non-topological VP ferromagnets. In our work, we utilize a continuum model that directly inherits its parameters from first-principles DFT calculations, making our results more directly connected to the material's underlying physics. While many of the magnetic and topological phases identified here are qualitatively similar to those found in previous studies, the displacement-field-induced phase transitions and the unique physics at larger twist angles observed in our results have not been previously explored. Moreover, by analyzing the magnon spectrum, we find that topological ferromagnets are considerably more robust against thermal fluctuations compared to their non-topological counterparts.

In this study, we have analyzed the different magnetic orders, their tunability, and thermal stability in twisted TMD. 
Our results make several testable predictions for future experimental studies.
By increasing the vertical displacement field, it is possible to sequentially transition from the topological VP state to the trivial VP state, and finally to the IVC state, as depicted in Fig.~\ref{fig:cant}. 
A displacement field tuned transition from VP to a non-VP phase has recently been observed~\cite{Xiaodong} consistent with our theory, and our analyses reveal several new aspects.
Namely, we predict that this progression involves two key transitions. The first topological transition from the topological VP state to the trivial VP state can be investigated using transport experiments that measure the quantized Hall conductance. 
The second soft canting transition from the trivial VP state to the IVC state can be directly observed optically via magnetic circular dichroism (MCD), which probes the degree of valley polarization. To distinguish the IVC state from other valley-unpolarized states, which could either be metallic or exhibit topological order, one can measure the magnon propagation. Since the IVC state behaves as a magnon superfluid, with magnons being linearly dispersed as shown in Fig.~\ref{fig:magnon} (a), the magnon wavepacket can propagate at a constant velocity without spreading out \cite{Chenhao}.

Our theory demonstrates that the ferromagnetic ordering temperature is strongly influenced by the topology of the electronic states. In the topologically trivial VP phase, long-range ferromagnetic order is suppressed at finite temperatures due to an approximate $SU(2)$ symmetry. One concrete prediction is that the ferromagnetic ordering temperature, as indicated by magnetic hysteresis, will be much lower than the Curie-Weiss temperature derived from high-temperature magnetic susceptibility fits.

Physically, this discrepancy arises because, between the ferromagnetic ordering temperature and the Curie-Weiss temperature, the system only forms ferromagnetic domains, with each domain having magnetization pointing in opposite direction. From the magnon perspective, low-wavelength magnons with energies below this temperature proliferate, which destroys ferromagnetism at long-range and favors domain formation instead. As a result, despite the charge gap in twisted homobilayer TMDs being similar to that in twisted bilayer graphene (both around \SI{30}{K}), the temperature at which Hall conductance quantizes is much higher in TMDs \cite{Serlin2019,FQAH}. Here the Hall conductance serves as an accurate probe of long-range ferromagnetism since it changes dramatically when domains disappear and polarization becomes impossible.

This highlights the physical significance of the magnon spectrum: despite the presence of a large charge gap, the magnetic ordering temperatures in TMDs are lower, as they are governed by the magnon gap. Thus, our findings provide valuable insights for guiding the search for highly tunable and robust ferromagnetic states in the moiré platform.

{\em Note added.} After we post this manuscript, we become aware of a preprint that studies twisted \ce{MoTe2} at $\theta = 3.7 \degree$ using self-consistent Hartree Fock and density matrix renormalization group (DMRG) numerics, which agrees with our phase diagram at $\theta = 3.7 \degree$ \cite{AshvinCFL}.

\begin{acknowledgments}
We thank Xiaodong Xu, Shubhayu Chatterjee, Tomohiro Soejima, Daniel Parker, and Minxuan Wang for helpful discussions. We thank Nick Bultinck and Tomohiro Soejima for their contributions to the code used to perform the HF and TDHF calculations. We thank Yang Zhang for providing the TMD atomic structure figure. This work was supported by the Air Force Office of Scientific Research (AFOSR) under award FA9550-22-1-0432 (TD and LF), and  the U.S. Department of Energy, Office of Science, Office of Basic Energy Sciences, Materials Sciences and Engineering Division under Contract No. DE-AC02-05-CH11231 (van der Waals Heterostructures Program KCWF16 and Theory of Materials program KC2301) (TW and MZ). 
This research used the Lawrencium computational cluster resource provided by the IT Division at the Lawrence Berkeley National Laboratory (Supported by the Director, Office of Science, Office of Basic Energy Sciences, of the U.S. Department of Energy under Contract No. DE-AC02-05CH11231).

\end{acknowledgments}

\bibliography{mainPRL.bib}

\newpage

\onecolumngrid

\vspace{0.3cm}

\supplementarysection
 
\begin{center}
\Large{\bf Supplemental Material: Magnetic orders and quantum anomalous Hall effect\\ in twisted bilayer \ce{MoTe2} and \ce{WSe2}}
\end{center}

\section{Continuum model of twisted TMDs}

In the main text, we write down the continuum model Hamiltonian of twisted TMD,
\begin{equation} 
\begin{gathered}
    H_0=\sum_{\tau, l, \mathbf{r}} c_{\tau, l, \mathbf{r}}^{\dagger}\left([h_{\tau}]_{ll'}-\mu \delta_{l l'}\right) c_{\tau, l', \mathbf{r}}, \quad
    [h_{K}]_{ll'}=\left(\begin{array}{cc}h^K_+ + V_{+}(\boldsymbol{r}) - u_D & T(\boldsymbol{r}) \\ T^{\dagger}(\boldsymbol{r}) & h^K_- + V_{-}(\boldsymbol{r}) + u_D\end{array}\right)
\end{gathered}
\end{equation}
where $\tau$ is the valley index and $l$ is the layer index. The kinetic part $h^K_{\pm}$ takes the form
\begin{equation} \label{eq:potential}
    h^K_{\pm} = -\frac{\hbar^2\left(-i \nabla-\kappa_{\pm}\right)^2}{2 m^*}
\end{equation}
where $m^*$ is the effective mass. Here $\kappa_{+}$ and $\kappa_{-}$ are the $K$ points of the top and bottom layer respectively, which got folded to the moir\'e Brillouin zone corners (see Fig.~2 (a)). The moir\'e structure results in an intralayer moir\'e potential,
\begin{equation}
    V_{\pm}(\mathbf{r})=2 V \sum_{j=1,3,5} \cos \left(\mathbf{g}_j \cdot \mathbf{r} \pm \phi \right)
\end{equation}
and an interlayer tunneling,
\begin{equation}
    T(\mathbf{r})=w\left(1+e^{-i \mathbf{g}_2 \cdot \mathbf{r}}+e^{-i \mathbf{g}_3 \cdot \mathbf{r}}\right)
\end{equation}
where $\mathbf{g}_j =\frac{4 \pi}{\sqrt{3} a_M}(\cos \frac{\pi(j-1)}3, \sin \frac{\pi(j-1)}3)$ are the moir\'e reciprocal lattice vectors. Here $a_M=\frac{a_0}{2 \sin (\theta / 2)}$ is the moir\'e lattice constant, with $\theta$ being the twist angle and $a_0$ being the atomic lattice constant. Here we write down the $K$ valley Hamiltonian $h_K$ explicitly, while $h_{K'}$ can be obtained as the time-reversal conjugate of $h_K$. 
In this work, we use continuum model parameters obtained from recent large scale DFT calculations, see Table~\ref{tab:parameter} \cite{LiangNC,LiangFQH}.

\begin{table}[!htbp]
    \centering
    \renewcommand*{\arraystretch}{1.2}
    \begin{tabular}{cccccc}
    \hline \hline Materials & $\phi \;(\degree)$ & $V \;$(meV) & $w \;$(meV) & $m^* \;(m_e)$ & $a_0 \;(\AA)$ \\
    \hline \ce{MoTe2} & $-91$ & $  11.2$ & $-13.3$ & $0.62$ & $3.47$ \\
    \ce{WSe2} & $128$ & $9$ & $18$ & $0.43$ & $3.32$ \\
    \hline \hline
    \end{tabular}
    \caption{Continuum model parameters of twisted TMDs.}
    \label{tab:parameter}
\end{table}

\begin{figure}[!htbp]
    \centering
    \includegraphics[width = 0.48\textwidth]{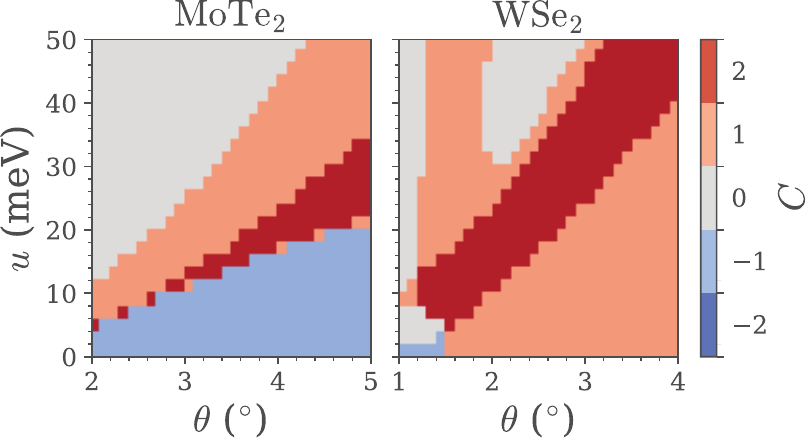}
    \caption{Valley Chern number of the \textit{second} valence band in the $K$ valley as a function of twist angle and displacement field.}
    \label{fig:chern2}    
\end{figure}

In the main text, we present the bandwidth and the valley Chern number of the first valence band (see Fig.~\ref{fig:free} (c)). Now we will briefly discuss the valley Chern number of the second valence band. As shown in Fig.~\ref{fig:chern2}, the valley Chern number has a complicated dependence on both the twist angle $\theta$ and the displacement field $u_D$. We note that at parameter regime where the first valence band is topological, the valley Chern number of the second valence band is the same as the first valence band. Therefore, simply including both bands does not admit any tight-binding description in the parameter regime of interest. In the interaction-driven trivial VP phase, the second valence band usually acquires higher Chern number so that the filled first valence band can be trivial.

\section{Interacting Hamiltonian and interaction-driven phases}

\subsection{Interacting Hamiltonian and self-consistent Hartree-Fock (HF) numerics}

Now we introduce the interacting Hamiltonian,
\begin{equation}
    H = H_0 + H_C, \quad H_C=\frac{1}{2 A} \sum_{\mathbf{q}} V_C(\mathbf{q}): \rho(\mathbf{q}) \rho(-\mathbf{q}):
\end{equation}
where $A$ is the sample area, $V_C(\mathbf{q})=e^2 \tanh (q D) /(2 \epsilon q)$ is the repulsive dual gate screened Coulomb interaction with sample-gate distance $D = \SI{30}{nm}$, and $\rho(\mathbf{q})$ is the Fourier transform of the electron density operator $\rho(\mathbf{r})=\sum_{\tau, l, \mathbf{r}} c_{\tau, l, \mathbf{r}}^{\dagger} c_{\tau, l, \mathbf{r}}$. Here we neglect any intervalley Hund's coupling that could either come from Coulomb interaction or phonon-induced coupling. We emphasize that the magnetic anisotropy we focus on in this work is induced by the kinetic energy, even in the absence of any intervalley Hund's coupling. For future convenience, we introduce the moir\'e Bloch electron operator $\psi_{n, \tau, \mathbf{k}}^{\dagger} = \sum_l u^*_{n, \tau, l, \mathbf{k}} c_{\tau, l, \mathbf{k}}^{\dagger}$ with $u_{n, \tau, l, \mathbf{k}}^*$ being the Bloch wave functions and $n$ being the band index. The $\psi_{\tau, \mathbf{k}}^{\dagger}$ operator introduced in the main text is defined to sum over all bands, $\psi_{\tau, \mathbf{k}}^{\dagger} \equiv \sum_n \psi_{n, \tau, \mathbf{k}}^{\dagger}$. Now we can also write the density operator in terms of the intravalley form factor $\rho(\mathbf{q}) = \sum_{\tau, \mathbf{k},nn'} [\lambda^{\tau \tau}_{\mathbf{q}}(\mathbf{k})]^{n'n} \psi^\dagger_{n,\tau, \mathbf{k}} \psi_{n^\prime,\tau, \mathbf{k} + \mathbf{q}}$ where $[\lambda^{\tau' \tau}_{\mathbf{q}}(\mathbf{k})]^{n'n} = \bra{u_{n,\tau,\mathbf{k}}} \ket{u_{n^\prime, \tau^\prime, \mathbf{k} + \mathbf{q}}}$.

We study this interacting Hamiltonian using a self-consistent Hartree-Fock (HF) calculation. To reduce the computational cost, we project the Coulomb interaction to the first two valence bands per valley, which allows us to use a $30 \times 30$ momentum grid to efficiently sweep the phase diagram and a $50 \times 50$ momentum grid to extract HF band structures. Keeping at least two bands per valley is essential at filling one hole per moir\'e unit cell since it allows an interaction-induced topological transition by mixing multiple bands, which plays an essential role in our study of magnetic orders and their magnetic anisotropy. Later, we will also justify that keeping two bands is sufficient for most quantities of interest. We note that the projected Hamiltonian does not admit any tight-binding description even after involving multiple bands since the net valley Chern number of these bands is still not necessarily vanishing.

Using the formulation described in Ref.~\onlinecite{NickPRX}, we solve the self-consistent equations for Slater determinant states characterized by the one-electron covariance matrix,
\begin{equation} \label{eq:P}
    [P^{\tau' \tau}_{\mathbf{Q}}]^{n' n}(\mathbf{k}) = \begin{cases} \langle\psi_{n, \tau, \mathbf{k}}^{\dagger} \psi_{n', \tau^{\prime}, \mathbf{k}}\rangle, \quad &\tau = \tau' \\
    \langle\psi_{n, \tau, \mathbf{k}}^{\dagger} \psi_{n', -\tau^{\prime}, \mathbf{k} + \mathbf{Q}}\rangle, \quad &\tau = - \tau' 
    \end{cases}
\end{equation}
where we allow a shift $\mathbf{Q}$ for any intervalley coherence \cite{Ben}. Then we use both the ODA and the EDIIS algorithms to solve the self-consistency equation \cite{ODA,EDIIS}. In general, we should allow all possible $\mathbf{Q}$ at the same time $P[\{\mathbf{Q}\}]$, but in this work we will focus on those states characterized by a single $\mathbf{Q}_0$, i.e. $P[\{\mathbf{Q}\}] = \delta_{\mathbf{Q},\mathbf{Q}_0}P_{\mathbf{Q}_0}$. We always find the lowest energy ground state to be realized when $\mathbf{Q}$ coincides with high-symmetry points $\mathbf{Q} = \mathbf{0}$ and $\mathbf{Q} = \kappa_-$ in the parameter space we study. We note that $\mathbf{Q} = \kappa_+$ and $\mathbf{Q} = \kappa_-$ are not equivlent since the displacement field breaks the $C_{2y}$ rotation symmetry that relates the two.

\begin{figure}[!htbp]
    \centering
    \includegraphics[width = 0.6\textwidth]{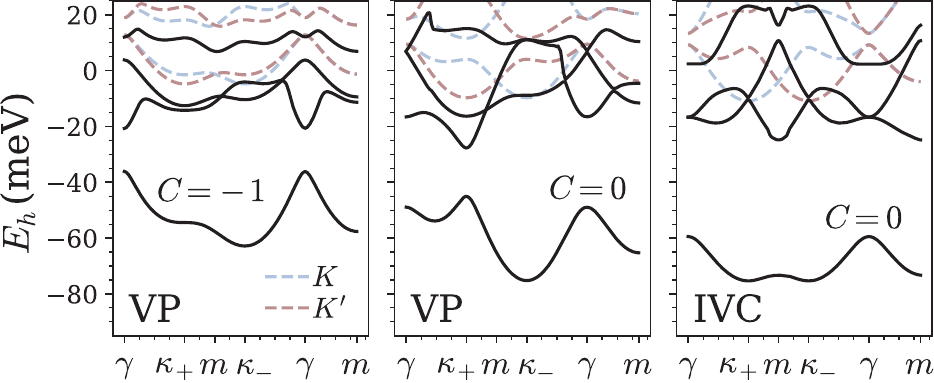}
    \caption{Typical HF band structure of twisted TMD in the topological VP, trivial VP and the IVC phase. We take twisted \ce{MoTe2} with $\theta = 3.5 \degree$ at $u_D = \SI{3}{meV}$ ($C = -1$ VP), \SI{9}{meV} ($C = 0$ VP), and \SI{20}{meV} (IVC$_-$) as examples. Here we show the hole energy spectrum, so the electron valence bands in Fig.~1 (b) becomes hole conduction bands here. We label the Chern number of the first HF hole conduction band. We overlay the free hole band structure of both valleys at the corresponding parameters in dashed lines. For VP phases we always plot the HF band structure when the state polarizes to the $K$ valley.}
    \label{fig:hf}    
\end{figure}

\subsection{Interaction-driven phases and their HF band structures}

\begin{table}[!htbp]
    \renewcommand*{\arraystretch}{1.3}
    \centering
    \begin{tabular}{ccccc}
    \hline \hline & & \multicolumn{2}{c}{Symmetry} & Topology\\ \cline{3-5} Magnetic orders & Valley-ordered phases & $\tilde{\mathcal{T}}$ & $U(1)_v$ & $C$\\
    \hline / & SM & $\checkmark$ & $\checkmark$ & 0 \\ QAH & topological VP & $\times$ & $\checkmark$ & $\pm 1$\\
    FM$_z$ & trivial VP & $\times$ & $\checkmark$ & $0$\\
    FM$_{xy}$ & IVC$_{\mathbf{0}}$ & $\checkmark$ & $\times$ & $0$\\
    $120 \degree$ AFM & IVC$_-$ & $\checkmark$ & $\times$ & $0$\\
    \hline \hline
    \end{tabular}
    \caption{Symmetry and topology of various valley-ordered states. Here $\tilde{\mathcal{T}}$ is time-reversal combined with $180 \degree$ spin rotation.
    }
    \label{tab:symmetry}
\end{table}

\begin{figure}[htbp]
    \centering
    \includegraphics[width = 0.48\textwidth]{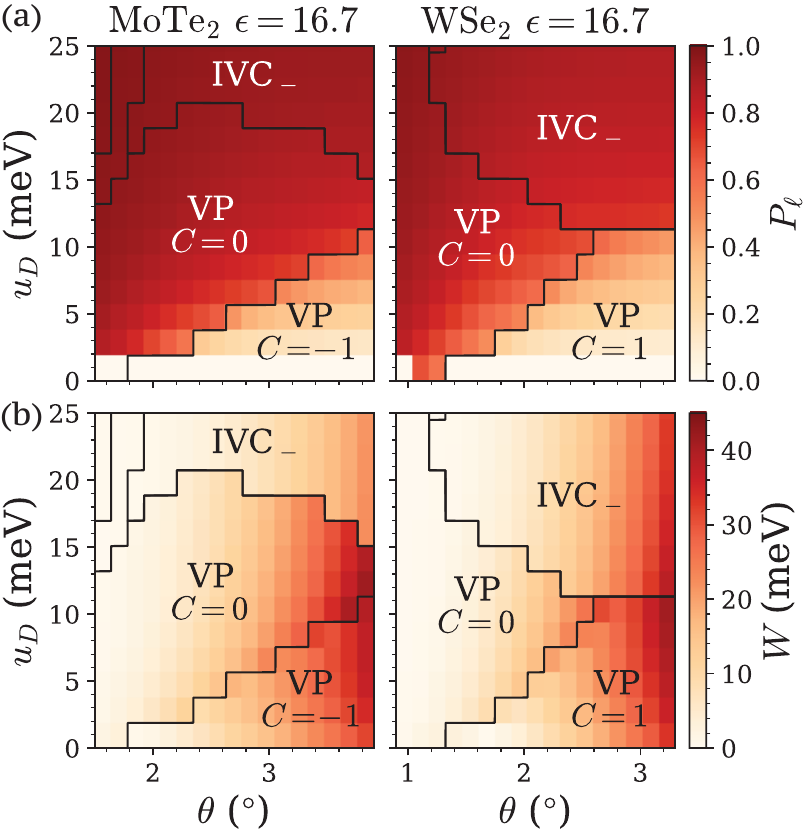}
    \caption{Interacting phase diagram of twisted \ce{MoTe2} and \ce{WSe2} as a function of displacement field and twist angle. We show the renormalized band width $W$ in (a) and the layer polarization $P_l$ in (b). Here $P_l \equiv (n_t - n_{b})/(n_t + n_{b})$, $n_l \sim \sum_{\mathbf{k}} \langle c_{l,\mathbf{k}}^{\dagger}(\boldsymbol{k}) c_{l,\mathbf{k}}(\boldsymbol{k})\rangle$ is the occupation of the $l$ layer, and $c_{l,\mathbf{k}}^{\dagger} \equiv \sum_{\tau} c_{\tau, l,\mathbf{k}}^{\dagger}$.}
    \label{fig:layer}    
\end{figure}

We start with how we identify different phases in HF. Depending on which symmetries are explicitly enforced, we can find three self-consistent solutions: (i) a valley polarized (VP) state which breaks the $\tilde{\mathcal{T}} = \mathcal{T} e^{i \pi s^z} $ symmetry (time-reversal combined with $180 \degree$ spin rotation around the $z$ axis~\cite{LiangPRX}) but preserves the valley $U(1)_v$ symmetry, (ii) a inter-valley coherent (IVC) state which breaks $U(1)_v$ symmetry but preserves the $\tilde{\mathcal{T}}$ symmetry, and (iii) a fully symmetric metal (SM) that respects both of the symmetries (see Table~\ref{tab:symmetry}). 
When we study transition between VP and IVC, we allow both $\tilde{\mathcal{T}}$ and $U(1)_v$ breaking, which is necessary to capture the smooth canting transition from trivial VP to IVC.
We only find SM to be favorable under an unrealistically large dielectric $\epsilon > 50$ in the parameter regime of interest. We further distinguish the topological VP phase and the trivial VP phase by computing the Chern number of the filled first HF valence band. The momentum shift $\mathbf{Q}$ in the intervalley sector is irrelevant to the VP phase whose covariance matrix only has intravalley components. However, IVC state with different momentum shift $\mathbf{Q}$ can have different energy. We always find IVC state with momentum shift $\mathbf{Q} = \mathbf{0}$ and $\mathbf{Q} = \kappa_-$ to be most favorable, whose competition further determines the phase diagram within the IVC phase. 

We show Hartree-Fock band structures of twisted \ce{MoTe2} at $\theta = 3.5 \degree$ in Fig.~\ref{fig:hf} as an example, where we take the same parameters as the soft mode band structures in Fig.~4. For the VP state, the first HF band follows the first non-interacting band of the valley it polarizes to ($K$ valley in Fig.~\ref{fig:hf} (a)); while for the IVC state, it hybridizes the first non-interacting band from two valleys. We note a significant magnification of both the band gap and the band width after we introduce interaction. In addition, we also see that the band gap is slightly larger and the band width is narrower in the IVC state since it involves only the lower energy portion of the first band. These features are most obvious in the charge gap plot in Fig.~3 (a) and the renormalized band width plot in Fig.~\ref{fig:layer} (a). We find a sudden drop of the band width across the VP to IVC transition.

The band structure of the topological VP state and the trivial VP state differ in a more subtle way. Compared to the topological VP state, the trivial VP state deviates from the non-interacting band structure at the $\kappa_+$ point (see Fig.~\ref{fig:hf}). To understand this behavior, we note that the trivial VP state is closely associated with spontaneous layer polarization (see Fig.~\ref{fig:layer} (b)) \cite{LiangNC,Multiferroicity}. 
Interactions will tend to increase layer polarization, and the layer polarized limit is necessarily trivial (in the tight binding limit~\cite{LiangNC}, this is understood as an increased sublattice polarization due to nearest neighbor interactions).
Even though the displacement field explicitly breaks the $C_{2y}$ symmetry that swaps two layers, we still find a sudden increase in layer polarization across the topological to trivial transition up to $u_D = \SI{10}{meV}$. At small angles $\theta \lesssim 1.6 \degree$, we find spontaneous layer polarization down to zero field.

\begin{figure}[htbp]
    \centering
    \includegraphics[width = 0.75\textwidth]{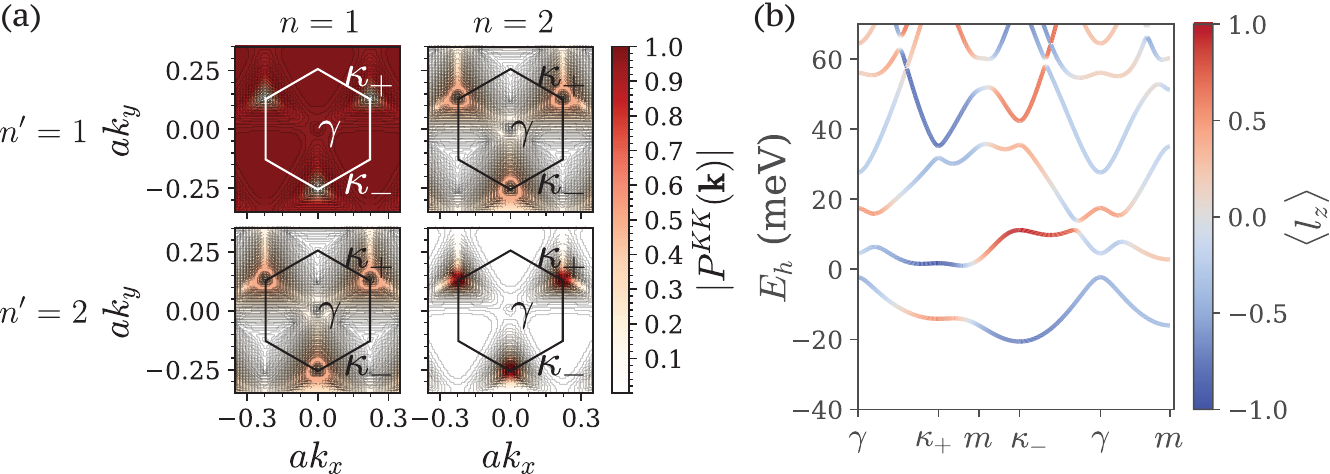}
    \caption{(a) Nontrivial elements of the one-electron covariance matrix $[P^{\tau' \tau}_{\mathbf{0}}]^{n' n}(\mathbf{k}) =  \delta_{\tau' K} \delta_{\tau K} [P^{KK}_{\mathbf{0}}]^{n n^{\prime}}(\mathbf{k})$ in the trivial VP phase (polarized to the $K$ valley). Here we take twisted \ce{MoTe2} at $\theta = 3.5 \degree$, $u_D = \SI{11}{meV}$ and $\epsilon = 16.7$ in both (a) and (b).  (b) Layer polarization of the non-interacting band structure in the $K$ valley.}
    \label{fig:VP}    
\end{figure}

With the layer polarization picture in mind, we can examine how the interaction mixes the first few bands in detail. As shown in Fig.~\ref{fig:VP} (b), the $\kappa_+$ point of the first noninteracting band mainly comes from the top layer, while the rest of the band mainly comes from the bottom layer. On the other hand, the $\kappa_+$ point of the second band comes from the bottom layer. To polarize the state to the bottom layer, the interaction must mix the first two bands around $\kappa_+$. Indeed, as shown in Fig.~\ref{fig:VP} (a), the covariance matrix $P^{KK}$ has the highest weight in the first band, except for the $\kappa_+$ point where the weight is shifted to the second band. 
Furthermore, the vanishing residue weight on the third band also indicates that it is sufficient to keep only the first two noninteracting bands in this parameter regime.

\subsection{Convergence of the HF calculation}

\subsubsection{Convergence in the momentum cutoff of the interaction}

\begin{figure}[htbp]
    \centering
    \includegraphics[width = 0.4\textwidth]{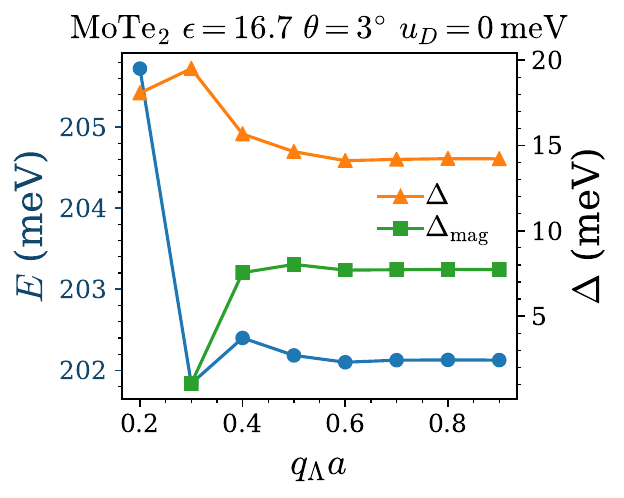}
    \caption{Energy per particle \( E \) (left axis) and charge gap \( \Delta \), magnon gap \( \Delta_{\text{mag}} \) (right axis) as functions of the momentum cutoff \( q_{\Lambda} a \) for \(\ce{MoTe2}\) at a typical twist angle \( \theta = 3^\circ \), dielectric constant \( \epsilon = 16.7 \), and displacement field \( u_D = 0 \, \text{meV} \).}
    \label{fig:qcut}    
\end{figure}

In Hartree-Fock (HF) calculations, we limit the interaction \( V_C(\mathbf{q}) \) to a finite range \( |\mathbf{q}| < q_{\Lambda} \) primarily for physical and practical reasons. From a physical perspective, keeping \( q_{\Lambda} \) smaller than the intervalley momentum transfer \( |\mathbf{K}| \) avoids introducing intervalley Hund’s coupling, which would otherwise break the approximate \( U(1) \times U(1) \) symmetry between the two valleys. Additionally, restricting the interaction to a finite range in momentum space accelerates the convergence of the HF optimization and avoids many local minimums in the optimization manifold. A further operational benefit is that this restriction reduces computational cost by limiting the number of form factors \( \lambda_{\mathbf{q}}^{\tau \tau'}(\mathbf{k}) \) that need to be computed.

However, \( q_{\Lambda} \) must also be large enough to properly capture the short-range part of the interaction energy. In Fig.~\ref{fig:qcut}, we illustrate the dependence of HF convergence on \( q_{\Lambda} \), examining several physical observables. For instance, at a twist angle of \( \theta = 3^\circ \), convergence is achieved for \( q_{\Lambda} a > 0.7 \). It is important to highlight that the required value of \( q_{\Lambda} \) for convergence depends sensitively on the twist angle \( \theta \). Empirically, \( q_{\Lambda} \) needs to exceed three times the moiré reciprocal lattice vector \( |\mathbf{g}| \) for reliable results. In practice, we adopt a \( \theta \)-dependent \( q_{\Lambda} \) to strike a balance between computational efficiency and convergence.

\subsubsection{Convergence in the number of the active bands}

\begin{figure}[htbp]
    \centering
    \includegraphics[width = 0.4\textwidth]{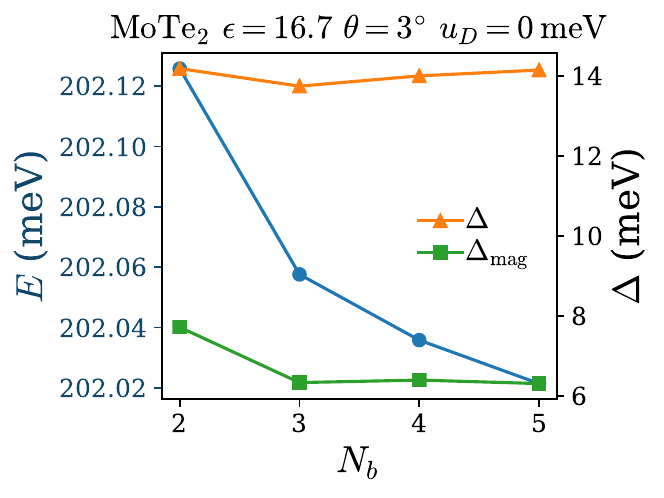}
    \caption{Energy per particle \( E \) (left axis) and charge gap \( \Delta \), magnon gap \( \Delta_{\text{mag}} \) (right axis) as functions of the number of active bands per valley \( n_b \) for \(\ce{MoTe2}\) at a typical twist angle \( \theta = 3^\circ \), dielectric constant \( \epsilon = 16.7 \), and displacement field \( u_D = 0 \, \text{meV} \).}
    \label{fig:nb}    
\end{figure}

\begin{figure}[htbp]
    \centering
    \includegraphics[width = 0.75\textwidth]{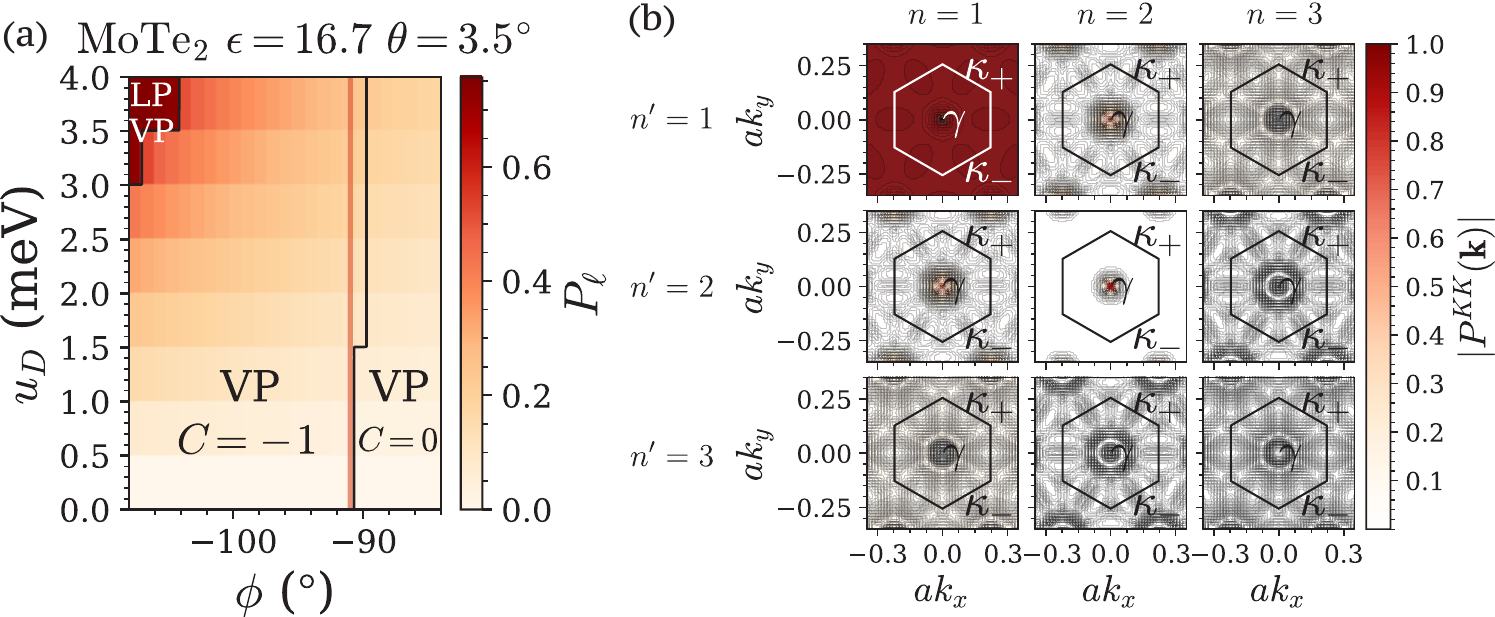}
    \caption{(a) Interacting phase diagram of twisted \ce{MoTe2} after including three \textit{active} bands per valley, shown as a function of displacement field $u_D$ and the phase $\phi$ of the moir\'e potential. We show the layer polarization $P_{\ell}$ and label the regime where the ground state exhibits a non-zero Chern number. The LP VP $C = 0$ phase corresponds to the trivial VP phase discussed in the main text, while the VP $C = 0$ phase is a newly identified phase discussed in the Supplementary Material. The red line cut indicates the value of $\phi$ used in the main text.
    (b) Nontrivial elements of the one-electron covariance matrix $[P^{\tau' \tau}_{\mathbf{0}}]^{n' n}(\mathbf{k}) = \delta_{\tau' K} \delta_{\tau K} [P^{KK}_{\mathbf{0}}]^{n n'}(\mathbf{k})$ in the layer-unpolarized trivial VP phase (polarized to the $K$ valley). Here, we use parameters for twisted \ce{MoTe2} at $\theta = 3.5^\circ$, $u_D = \SI{0}{meV}$, and $\epsilon = 16.7$. The phase $\phi = 88^\circ$ is manually adjusted to stabilize the layer-unpolarized trivial VP state.}
    \label{fig:nLP}    
\end{figure}

The convergence with respect to the number of active bands in Hartree-Fock (HF) calculations requires careful consideration. Including more bands allows states to lower their energy by accessing higher bands, which is generally beneficial as it suggests the system is approaching the correct physical Hilbert space. As shown in Fig.~\ref{fig:nb}, states can indeed reduce their energy when more active bands are included. For states discussed in the main text, the energy reduction remains within $0.1\%$, leaving no changes in the phase diagram. Second, adding more bands introduces new degrees of freedom, which can lead to new, potentially competing states. However, whether these new degrees of freedom are physically relevant depends on the system under study.

In the case of twisted \ce{MoTe2} and \ce{WSe2}, adding a third band in the HF calculation may introduce degrees of freedom that are not physically accurate, due to discrepancies between the continuum model and DFT results. The parameters of the continuum model are specifically fitted to describe the first two bands \cite{LiangFQH}, and significant deviations from DFT begin to appear when the third band is included. Thus, including the third band may not reflect the correct physics of the system.

When the third band is included in HF calculations, a competing state can emerge due to the availability of additional degrees of freedom. Specifically, a layer-unpolarized trivial VP state arises in twisted \ce{MoTe2}. This state differs from the trivial VP state discussed in the main text, which is always spontaneously layer-polarized. We show the covariance matrix $P^{KK}$ of the layer-unpolarized trivial VP state in Fig.~\ref{fig:nLP}(b), which clearly contrasts with the layer-polarized trivial VP (LP VP) state shown in Fig.~\ref{fig:VP}(a). In this case, the interaction primarily couples the first two bands at the $\gamma$ point, rather than at the $\kappa_+$ point. The small but nonzero weight on the third band indicates that this state becomes favorable only when three bands are included in the HF calculation. If we truncate the covariance matrix to the first two bands and renormalize it, this state would have a much higher energy than the topological VP state.

The competition between the layer-unpolarized trivial VP state and the topological VP state depends sensitively on the parameters of the continuum model, particularly the phase \( \phi \) of the moir\'e potential in Eqn.~(\ref{eq:potential}). As shown in Fig.~\ref{fig:nLP}(a), the topological VP phase is favored for $|\phi| \gtrsim 91^\circ$, which corresponds to the current value of the continuum model parameter and lies well within the fitting error of large-scale DFT calculations \cite{LiangNC, LiangFQH}. Furthermore, the topological VP (QAH) phase has already been observed experimentally in twisted \ce{MoTe2} at $\theta = 3.5^\circ$ and $\epsilon \sim 15$, helping to pinpoint the phase $\phi$ of the moir\'e potential \cite{Xiaodong, XiaodongFQH}. In the main text, we limit our HF calculations to two active bands to avoid the introduction of spurious competing states that may not be physically meaningful.

\section{Soft modes in the VP and the IVC phase}

To understand low-energy excitations in these interaction-driven phases, which determine their thermal stability, we use the time-dependent Hartree-Fock calculation to study the eigenmodes of these low-energy excitations and their energetics \cite{wu2020ferromagnetism,wu2020collective,kwan2021exciton}. Following the formulation described in Ref.~\onlinecite{SoftMode}, the starting point is our HF ground state characterized by the covariance matrix $P(\mathbf{k})$. We first write down the following ansatz for low energy bosonic excitations characterized by the wavefunction $[\phi_{\boldsymbol{q}}^{\tau' \tau}]^{n'n}(\boldsymbol{k})$, 
\begin{equation}
    \hat{\phi}_{\boldsymbol{q}}=\sum_{\boldsymbol{k}} \psi_{n, \tau, \mathbf{k}}^{\dagger} [\phi_{\boldsymbol{q}}^{\tau' \tau}]^{n'n}(\boldsymbol{k}) \psi_{n', \tau^{\prime}, \mathbf{k}+\mathbf{q}}
\end{equation}
Now we look for the eigenmodes satisfying
\begin{equation}
    \left \langle \left[\hat{H},\hat{\phi}_{\boldsymbol{q}}\right] \right \rangle_{\mathrm{MF}}=\omega_{\boldsymbol{q}} \hat{\phi}_{\boldsymbol{q}}
\end{equation}
where $\hat{H}$ is the full interacting Hamiltonian, and $\langle \cdot \rangle_{\mathrm{MF}}$ means performing partial Wick contractions with the HF ground state $P(\mathbf{k})$ to reduce all four-fermion operators to two-fermion operators. The eigenmodes can be computed by diagonalizing a quadratic boson Hamiltonian (see Ref.~\onlinecite{SoftMode} App. A for details). Due to convergence challenges, we restrict ourselves to a $26 \times 26$ momentum grid for the gapless Goldstone modes in the IVC phase.

\begin{figure}[!htbp]
    \centering
    \includegraphics[width = 0.46\textwidth]{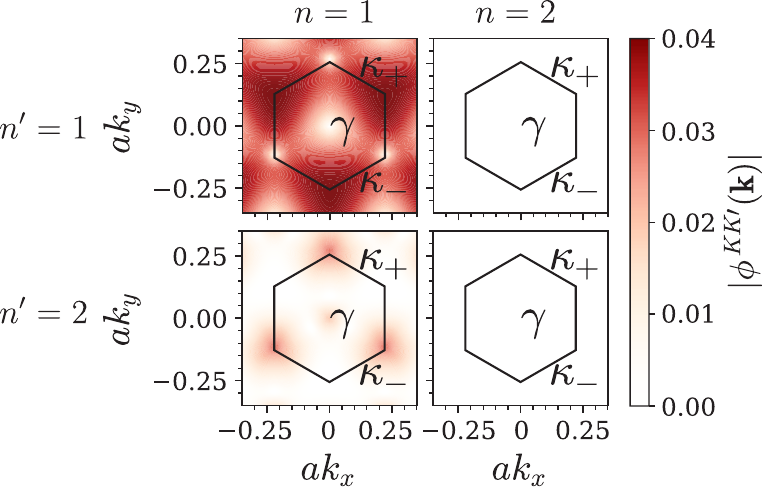}
    \caption{Nontrivial elements of the momentum $\mathbf{Q} = \kappa_-$ magnon wavefunction $[\phi^{\tau' \tau}_{\mathbf{\mathbf{Q}}}]^{n' n}(\mathbf{k}) =  \delta_{\tau' K} \delta_{\tau K'} [\phi^{KK'}_{\mathbf{Q}}]^{n n^{\prime}}(\mathbf{k})$ in the VP phase (polarized to the $K$ valley). Here we take twisted \ce{MoTe2} at $\theta = 3.5 \degree$ and $u_D = \SI{1.2}{meV}$ in the topological VP phase as an example.}
    \label{fig:magnonstate}
\end{figure}

To understand the magnon gap in the VP phase (topological or trivial), we can look at the nature of the soft mode at the soft mode bottom. For vast of the phase diagram, the soft mode band bottom appears at $\mathbf{Q} = \kappa_-$. We show the wavefunction $[\phi^{\tau' \tau}_{\mathbf{\mathbf{Q}}}]^{n' n}(\mathbf{k})$ of the momentum $\mathbf{Q}$ soft mode in Fig.~\ref{fig:magnonstate}. The most important character is that it only contains $\phi^{KK'}$ components, so it indeed corresponds to a magnon instead of an exciton. Condensing such a magnon will give you the IVC$_-$ state, which relates the magnon gap to the energy difference between the VP state and the IVC state.

\section{Non-linear Sigma Model Description of the Magnetic Order}

\subsection{Non-linear Sigma Model}

To analyze the magnetic orders and their associated collective excitations (magnons) in twisted TMD systems, we construct a non-linear sigma model. While this model resembles the one used for twisted bilayer graphene (TBG), in the case of twisted TMDs, the four isospin flavors reduce to two valleys, \( \tau = K, K' \). At the filling \( \nu_h = 1 \), the valley order is described by a traceless \( 2 \times 2 \) matrix \( Q \), defined as
\begin{equation}
    Q_{\alpha \beta}(\boldsymbol{k}) = \left\langle\left[\psi_{\alpha, \boldsymbol{k}}^{\dagger}, \psi_{\beta, \boldsymbol{k}}\right]\right\rangle,
\end{equation}
which relates to the projector \( P(\boldsymbol{k}) \) in Eq.~(\ref{eq:P}) through \( Q(\boldsymbol{k}) = 2 P(\boldsymbol{k}) - 1 \). This matrix captures the valley polarization in the system. For the valley-polarized (VP) state, \( Q(\boldsymbol{k}) \) is approximately constant and is denoted by \( \tilde{Q} \). For the inter-valley coherent (IVC) state, where hybridization between opposite Chern number bands occurs, \( Q_{\boldsymbol{k}} \) exhibits winding around the Brillouin zone. To describe the IVC state, we adopt the following ansatz:
\begin{equation}
    Q_{\boldsymbol{k}} = e^{i \varphi(\boldsymbol{k}) \tau_z} \tilde{Q} e^{-i \varphi(\boldsymbol{k}) \tau_z}
\end{equation}
where \( \varphi_{\boldsymbol{k}} = \arg(k_x + i k_y) \). Under this ansatz, the IVC order parameter \( \Delta(\boldsymbol{k}) = \left\langle\psi_{K, \mathbf{k}}^{\dagger} \psi_{K^{\prime}, \mathbf{k}}\right\rangle \) takes the form \( e^{2i \varphi_{\boldsymbol{k}}} \), up to a uniform phase. In the valley basis, the VP state is described by \( \tilde{Q}_{\mathrm{VP}} = \tau_z \), while the IVC state is represented by \( \tilde{Q}_{\mathrm{IVC}} = \cos \phi \tau_x + \sin \phi \tau_y \), reflecting a coherent superposition of the two valleys.

The non-linear sigma model for \( \tilde{Q} \) can now be constructed. The kinetic term governing the system dynamics is formulated as a one-dimensional Wess-Zumino term \cite{skyrmion}:
\begin{equation}
    \mathcal{L}[\tilde{Q}] = \frac{1}{2} \operatorname{tr} \left[ T(\boldsymbol{r})^{\dagger} \tilde{Q} \partial_t T(\boldsymbol{r}) \right] - \frac{\rho}{4} \operatorname{tr} \left[ \nabla \tilde{Q}(\boldsymbol{r}) \right]^2 - E[\tilde{Q}(\boldsymbol{r})]
\end{equation}
where \( \tilde{Q}(\boldsymbol{r}) = T^{\dagger}(\boldsymbol{r}) \tilde{Q} T(\boldsymbol{r}) \) allows for long-wavelength spatial variation. The energy term \( E[\tilde{Q}] \) encodes the anisotropies that break the valley \( SU(2) \) symmetry, which are crucial in determining the magnetic behavior of the system. Two specific anisotropy terms play a dominant role. The first is the interaction energy cost \( E_{\Delta} \) associated with the winding of the IVC order parameter. At the Hartree-Fock level,
\begin{equation}
    E_{\Delta} = \frac{1}{N} \sum_{\boldsymbol{k}} E_C(\boldsymbol{k})\left|\left(\nabla_{\boldsymbol{k}} - 2 i A_{\boldsymbol{k}}\right) \Delta(\boldsymbol{k})\right|^2
\end{equation}
We can rewrite \( E_{\Delta} = \frac{\alpha}{2} \operatorname{tr}\left( \tilde{Q} \tau_z \right)^2 \). The second important term is the kinetic energy gain \( E_h \) of the IVC state. In the coarse-grained nonlinear sigma model, we are unable to fully capture the canting of the isospin vector in momentum space, as it appears in a complicated form within the gradient term. However, we can partially account for this effect by integrating out these fluctuation modes, leading to \cite{skyrmion}
\begin{equation}
    E_h = - \frac{J}{4} \operatorname{tr} \left[\left( \tilde{Q} \tau_x \right)^2 + \left( \tilde{Q} \tau_y \right)^2 \right]
\end{equation}
These two terms give the non-linear sigma model as described in Eq.~(\ref{eq:nlsm}).

\subsection{Magnon Dispersion}

The magnon dispersions for both the VP and IVC states can be obtained by introducing a perturbation \( \delta \tilde{Q}(\mathbf{r}) \) to the uniform ground state. For the VP state, the perturbation takes the form
\begin{equation}
    \delta \tilde{Q}_{\mathrm{VP}}(\mathbf{r}) = \delta x(\mathbf{r}) \sigma_x + \delta y (\mathbf{r}) \sigma_y.
\end{equation}
Substituting this into the non-linear sigma model, we find the magnon dispersion for the VP state, $    \hbar \omega^{\mathrm{VP}}_{\mathbf{q}} = \rho q^2 + \alpha - J/2$. This dispersion is quadratic in \( q \), with the magnon gap determined by the anisotropy term \( \alpha \) and $J$.

For the IVC state, the low-energy excitations are phason modes, and the perturbation is given by
\begin{equation}
    \delta \tilde{Q}_{\mathrm{IVC}}(\mathbf{r}) = -\delta \phi(\mathbf{r}) \sin\phi\sigma_x + \delta \phi(\mathbf{r}) \cos\phi \sigma_y.
\end{equation}
In this case, we obtain a linearly dispersed, gapless Goldstone mode, which corresponds to the breaking of the valley \( U(1)_v \) symmetry, $\hbar \omega^{\mathrm{IVC}}_{\mathbf{q}} = \rho q$. 
This linear dispersion characterizes a gapless Goldstone mode due to the spontaneous breaking of \( U(1)_v \).

\vfill

\end{document}